\documentclass[12pt]{article}
\pdfoutput =1
\usepackage{float} %Include figure filesusepackage{graphicx} %Include figure files
\usepackage{amsmath}
\usepackage{slashed}
\usepackage{amsfonts}   
\usepackage{amssymb}

\begin{document}
\date{}
%%%%%%%%%%%%%%%%%%%%
\title{{\bf{\Large Broken Lifshitz invariance, spin waves and hydrodynamics}}}
%%%%%%%%%%%%%%%%%%%%
\author{
 {\bf {\normalsize Dibakar Roychowdhury}$
$\thanks{E-mail:  dibakarphys@gmail.com, dibakarr@iitk.ac.in}}\\
 {\normalsize  Indian Institute of Technology, Department of Physics,}\\
  {\normalsize Kanpur 208016, Uttar Pradesh, India}
\\[0.3cm]
}
%\date{}

\maketitle
%%%%%%%%%%%%%%%%%%%%%%%%%%%%%%%%%%%%%%%%%%%%%%%%%%%%%%%%%%%%%%%%%%%%%%%%%%%%
\begin{abstract}
In this paper, based on the basic principles of thermodynamics, we explore the hydrodynamic regime of interacting Lifshitz field theories in the presence of broken rotational invariance. We compute the entropy current and discover new dissipative effects those are consistent with the principle of local entropy production in the fluid. In our analysis, we consider both the parity even as well as the parity odd sector upto first order in the derivative expansion. Finally, we argue that the present construction of the paper could be systematically identified as that of the hydrodynamic description associated with \textit{spin waves} (away from the domain of quantum criticality) under certain limiting conditions.
\end{abstract}

%%%%%%%%%%%%%%%%%%%%%%%%%%%%%%%%%%%%
\section{Overview and Motivation}
Low temperature phases of various \textit{strange metal} systems have been a mysterious issue since their discovery. One of the key reasons for this lies behind the fact that unlike the ordinary metallic systems, strange metals do not posses any \textit{quasiparticle} description close to its Fermi surface. These systems are therefore known as \textit{Non Fermi liquids}\footnote{The enthusiastic reader is also referred to the review \cite{Iqbal:2011ae} and references therein.} \cite{QC1}-\cite{Si:2011hh}.

Matter in this quantum critical domain (so called \textit{quantum liquids}) exhibits certain fascinating features as the temperature is lowered substantially:\\
(1) The ratio between the specific heat to the temperature turns out to be very large indicating the fact that the effective mass for electrons could be larger many orders in magnitude than the free electron mass near its Fermi surface.\\
(2) The resistivity grows linearly with temperature namely, $ \rho \sim T $ which is also strange from the point of view of a Fermi liquid theory.\\
All the above features therefore strongly suggest the breakdown of the usual Fermi liquid theory in the domain of quantum criticality and as a result the low temperature physics of strongly correlated many body systems has been remained as a puzzle for the past couple of decades.

However, for various reasons, it is widely believed that the physics within this quantum critical domain is mostly governed by the obvious scaling symmetries of the underlying Quantum Critical Point (QCP). It is by now quite evident that the isometry group associated with QCPs is that of Lifshitz type namely the scaling symmetries are anisotropic at the fixed point \cite{Brynjolfsson:2010rx},
\begin{eqnarray}
\textbf{x}\rightarrow \lambda \textbf{x},~~t\rightarrow \lambda^{z}t
\end{eqnarray}
where, $ z $ is the dynamic critical exponent \cite{Kachru:2008yh}-\cite{Taylor:2015glc}.

The purpose of the present paper is however not to explore physics within this domain of quantum criticality, rather slightly away from it by explicitly breaking symmetries associated with the fixed point. In the following, we illustrate this in a bit detail. 

It turns out that QCPs with Lifshitz isometry group explicitly break Lorentz boost invariance and on the other hand preserve rotational symmetry as well as the translational invariance. The long wavelength dynamics associated with these fixed points has been explored extensively both from the perspective of a fluid description as well as the AdS/CFT duality \cite{Hoyos:2013eza}-\cite{Pang:2009wa}. Keeping this literature in mind, in the present paper, we are however interested in the question whether there exists any notion of hydrodynamics beyond the domain of quantum criticality and a possible systematic formulation of it based on the fundamental principles of Thermodynamics. In the following, we systematically quote the purpose of the present analysis:\\
(1)  We would like to explore the validity of the hydrodynamic description in a situation when the full Lifshitz algebra is broken down to some reduced symmetry group of it. It would be indeed quite interesting to explore what happens if we relax the condition of rotational symmetry and in particular how does it affect the parity odd transports of the theory. Naturally, it relaxes many of the previous constraints of the theory and allows additional tensor structures in the constitutive relations those were previously incompatible with the underlying symmetry of the system.\\
(2) The ultimate goal of our analysis is to write down an \textit{entropy current} \cite{Bhattacharyya:2014bha}-\cite{Banerjee:2012iz} in the presence of broken symmetry generators and explore how does it constraint the additional transports of the system those have emerged due to the symmetry breaking effect. In other words, we would like to explore whether there exist new dissipative effects (those are consistent with the principle of local entropy production) away from the domain of quantum criticality. In this regard, we consider both the parity even as well as the parity odd sector of the hydrodynamic constitutive relations and in particular explore how do the parity odd transports are affected due to the presence of anisotropy in the system. In this sense, the present analysis is valid for all of those fluid systems whose underlying isometry group comprises that of the broken generators of the Lifshitz algebra.

Two questions are quite obvious at this stage, (1) How do we introduce anisotropy in our system and, (2) What are the low lying physical d.o.f. of the system that one would like to interpret in spite of the fact that hydrodynamics is an \textit{effective} description and does not rely on the microscopic details of the system.

The anisotropy that we introduce in our analysis is in fact quite similar in spirit to that with the earlier analysis \cite{Gahramanov:2012wz}-\cite{Florkowski:2008ag}. However, as we shall see shortly, the interpretation as well as the motivation for introducing such anisotropy is completely different in spirit from that of the relativistic scenario. 

In our analysis, we introduce anisotropy by means of a four vector, $ \mathfrak{v}^{\mu} $ pointing along a specific direction. We associate a term $ \Delta(=P_{\perp}-P_{\parallel}) $ with this four vector which corresponds to the fact that the pressure along two different directions are different even in the rest frame of the fluid. With the introduction of this anisotropy, our next task turns out to be to write down all possible tensor structures those are possible to construct out of the basic variables of the fluid system under consideration, namely the fluid velocity ($ \mathfrak{u}^{\mu} $), temperature ($ T $), chemical potential ($ \mu_{c} $) and the anisotropy ($ \mathfrak{v}^{\mu} $). More specifically, one should be able to write down a number of tensor structures within the constitutive relations with arbitrary transports which could be fixed later by demanding the positivity of the so called entropy current. Moreover, it is to be noted that as the present theory is considered to be a \textit{slow} deformation ($ |\Delta| \ll 1 $) of the original Lifshitz hydrodynamics, therefore one must recover the known results of the Lifshitz theory as the fixed point is approached ($ \Delta \rightarrow 0 $).

As far as the d.o.f. are concerned, the answer is rather tricky. However, we claim that the present analysis could be used to model the hydrodynamic description associated with \textit{spin waves} \cite{Sachdev:2012dq}-\cite{qc1} in certain appropriate limit. The reason for this claim is the following: First of all, one should note that \textit{spin waves} could be thought of as being low lying excitation of certain material like dimer \textit{antiferromagnet} above its ground state. In other words, spin waves could be thought of as being the \textit{gapless}\footnote{This renders that the underlying theory is scale invariant.} excitation (\textit{Goldstone modes}) associated with the broken $ O(3) $ symmetry\footnote{At the level of hydrodynamics such symmetry breaking effects should be manifested in terms of the emergence of the pressure anisotropy in the constitutive relations of the conserved quantities.}. These excitation therefore clearly break the underlying rotational invariance of the system and they vanish as one approaches the domain of quantum crticality \cite{Sachdev:2012dq}-\cite{qc1}. In the following, we illustrate these points in detail. In particular, we try to explain what do we exactly mean by spin wave hydrodynamics. 

It turns out that, the classical hydrodynamic description of spin waves is valid as long as spin wave fluctuations are smaller than the correlation length \cite{Sachdev:2012dq},
\begin{eqnarray}
\xi \sim \frac{c}{T}\log \left( \frac{T}{|\mu_{c}-\mu_{QCP}|^{z \nu}}\right) 
\end{eqnarray}
where, $ \mu_{QCP} $ is the value of the chemical potetial exactly at the critical point and the energy of the spin wave excitation could be thought of $ \leq  c \xi^{-1}$. In other words,  we are still in the region where the effective correlation is still much greater than the low temperature thermal fluctuations of the system. Therefore the hydrodynamic description is still valid. Our ultimate goal would be to investigate whether it is at all possible to write down an entropy current under such circumstances.  

The organization of the rest of the paper is the following: In Section 2, we provide
the basic anisotropic construction for the most general charged dissipative fluids
in the presence of a global $ U(1) $ anomaly. In Section 3, we compute the entropy current both for the parity even as well as the parity odd sector of the theory. Finally, we conclude in Section 4.

%%%%%%%%%%%%%%%%%%%%%%%%%%%%%%%%%%%%
\section{Basics}
The purpose of the present calculation is to explore the consequences of \textit{broken} rotational invariance on chiral Lifshitz hydrodynamics \cite{Hoyos:2015lra}. However, to start with we first introduce the notion of chiral anomaly in the context of Lifshitz hydrodynamics as\footnote{Reader should not get confused the index $ a (=1,..~,N)$ as the index associated with internal symmetry (non abelian) generators. This index simply corresponds to the fact that to start with we have $ N $ number of anomalous $ U(1) $ charges in our system. However, in the subsequent analysis we would perform our computation for the simplest case with $ a=1 $. The corresponding result for multiple abelian charges could be generalized easily.}\cite{Hoyos:2015lra},
\begin{eqnarray}
\partial_{\mu}\mathfrak{T}^{\mu \nu}&=&\mathfrak{F}^{a \nu \sigma}\mathfrak{j}^{a}_{\sigma}\nonumber\\
\partial_{\mu}\mathfrak{j}^{a \mu}&=& \mathfrak{c}^{abc}\mathfrak{E}^{b}.\mathfrak{B}^{c}
\label{E1}
\end{eqnarray}
where, the anomaly coefficients are given by the rank $ 3 $ tensor $ \mathfrak{c}^{abc} $. Here, $ \mathfrak{j}^{a}_{ \mu} $s are the $ N $ anomalous $ U(1) $ currents and $ \mathfrak{F}^{a }_{\nu \sigma}(=\partial_{\nu}\mathfrak{A}^{a}_{\sigma}-\partial_{\sigma}\mathfrak{A}^{a}_{\nu}) $ are the $ N $ abelian field strength tensors of the theory. The electric as well as the magnetic field strengths are respectively given by,
\begin{eqnarray}
\mathfrak{E}^{a \mu}&=& \mathfrak{F}^{a \mu \sigma}\mathfrak{u}_{\sigma}\nonumber\\
\mathfrak{B}^{a \mu}&=&\frac{1}{2}\varepsilon^{\mu \nu\lambda\sigma}\mathfrak{u}_{\nu}\mathfrak{F}^{a}_{\lambda \sigma}.
\end{eqnarray}

We start our analysis by writing down the constitutive relations (compatible with the reduced symmetries of the system) both for the stress tensor as well the charge currents associated with the underlying hydrodynamic description. We choose to work in the so called Landau frame \cite{Hoyos:2015lra}. 

We first focus on the stress tensor part. It typically consists of the following two pieces\footnote{The reader should be aware of the fact that to start with Lifshitz field theories are QFTs for which $ z>1 $.  As a result such theories by no means could be connected to relativistic QFTs with Lorentz invariance (for which, $ z=1 $). In other words, Lifshitz QFTs do not possess a smooth Lorentzian limit and therefore it is not correct to interpret Lifshitz QFTs as \textit{non relativistic} QFTs simply because they are lack of Lorentz boost generators. On the other hand, because of the presence of the \textit{gapless} excitation in the theory, one might still hope to find some upper bound for the speed of particles in the theory which could be identified as the speed of light ($ c $) as well. However, one could switch over to the corresponding \textit{non relativistic} regime by taking appropriate ($ v^{i}/c $) limit which would render non relativistic QFTs with broken Galilean boost invariance \cite{Hoyos:2013eza}.},
\begin{eqnarray}
\mathfrak{T}^{\mu \nu}=\mathfrak{T}^{\mu \nu}_{(ideal)}+\mathfrak{T}^{\mu \nu}_{(D)}
\label{E3}
\end{eqnarray}
where, $ \mathfrak{T}^{\mu \nu}_{(ideal)} $ is the non dissipative (ideal) part of the stess tensor. On the other hand, $ \mathfrak{T}^{\mu \nu}_{(D)} $ is the dissipative piece of the stress tensor that includes all the \textit{first} order derivative expansions such that in the Landau frame \cite{Hoyos:2013eza},
\begin{eqnarray}
\mathfrak{u}_{\mu}\mathfrak{T}^{\mu \nu}_{(D)}=0.
\label{e5}
\end{eqnarray}

Since the full Lifshitz isometry group is now broken down to some reduced subset of it, therefore, to start with and in principle we are allowed to incorporate more generic tensor structures in the constitutive relation those were previously disallowed due to the presence of the underlying rotational invariance. Therefore in some sense the analysis of the present paper generalizes all the previous studies in the literature \cite{Hoyos:2013eza}-\cite{Hoyos:2015lra}. On top of it,  due to the presence of additional transports in the theory, the present analysis also turns out to be a bit more intricate compared to its relativistic counterpart. In the following we elaborate these issues into much more detail. 

The starting point of our analysis is quite straightforward. We write all possible distinct tensor structures those are allowed by the (reduced) symmetry of the system (as well as consistent with the Landau frame criterion) and we express our constitutive relations as a linear combination of each of these structures. The coefficients associated with these tensor structures are finally constrained by the principle of positivity of entropy production. Those transports which do not seem to be consistent with this physical constraint are set to be zero although they are allowed by the symmetry of the system. Keeping these facts in mind, the ideal part of the stress tensor could be formally expressed as\footnote{This form of the stress tensor is unique and is indeed consistent with the Landau frame criterion namely, $ \mathfrak{T}^{\mu \nu}\mathfrak{u}_{\nu}=-\epsilon \mathfrak{u}^{\mu} $ \cite{Hoyos:2013eza}. Moreover, here $ \mathfrak{g}_{\mu \nu} $ is the background metric with signature $ (-,+,+,+) $.},
\begin{eqnarray}
\mathfrak{T}^{\mu \nu}_{(ideal)}=\left( \epsilon + P_{\perp}\right)\mathfrak{u}^{\mu}\mathfrak{u}^{\nu}+P_{\perp} \mathfrak{g}^{\mu \nu}- \Delta \mathfrak{v}^{\mu}\mathfrak{v}^{\nu}
\label{E5}
\end{eqnarray}
where, $ \Delta = P_{\perp}-P_{\parallel} $ is the difference between the transverse and the longitudinal pressures\footnote{Note that the above choice (\ref{E5}) also stems from the fact that at the ideal level the fluid stress tensor must be diagonal.} \cite{Gahramanov:2012wz}-\cite{Florkowski:2008ag}. Here, $ \mathfrak{u}^{\mu} $ is the usual fluid velocity and $ \mathfrak{v}^{\mu}$ is the spacelike vector that defines the anisotropy in the system and is pointing along the longitudinal axis\footnote{The longitudinal axis defines the anisotropy in the system along the direction perpendicular to the reaction plane. $ P_{\parallel} $ is the pressure along this longitudinal axis, whereas on the other hand, $ P_{\perp} $ is the pressure along the direction perpendicular to this axis.} such that they are mutually orthogonal to each other namely\footnote{It turns out that the Landau frame criterion (\ref{e5}) and the velocity normalization (\ref{E6}) are intimately related to each other. Note that the Landau frame criterion (\ref{e5}) looks quite natural in the sense that it simply implies the fact that in the rest frame of the fluid the zeroth component of the stress tensor must be equal to the rest energy (density) of the fluid and this is the universal claim that should be valid for all classes of QFTs irrespective of their underlying symmetry group. Now it turns out that, given the stress tensor (\ref{E3}), the above criterion could be satisfied iff we choose the specific normalization (\ref{E6}) for the velocity field ($ \mathfrak{u}^{\mu} $) which therefore turns out to be a natural choice even for Lifshitz systems.}\cite{Gahramanov:2012wz}-\cite{Florkowski:2008ag},
\begin{eqnarray}
\mathfrak{u}^{\mu}\mathfrak{u}_{\mu}=-1,~~\mathfrak{v}^{\mu}\mathfrak{v}_{\mu}=1,~~
\mathfrak{u}^{\mu}\mathfrak{v}_{\mu}=0.
\label{E6}
\end{eqnarray}

 Considering the \textit{local} rest frame of the fluid,
\begin{eqnarray}
\mathfrak{u}^{\mu}=(1,0,0,0),~~\mathfrak{v}^{\mu}=(0,0,0,1)
\end{eqnarray}
it is indeed quite straightforward to show,
\begin{eqnarray}
\mathfrak{T}^{\mu \nu}_{(ideal)}= diag(\epsilon , P_{\perp}, P_{\perp}, P_{\parallel})
\end{eqnarray}
which is clearly consistent with the fact that the momentum density vanishes in the local rest frame of the ideal fluid.

Before we proceed further, one non trivial yet simple check is inevitable. We consider the Lifshitz Ward identity\footnote{Note that the Ward identity is always satisfied if the underlying scaling symmetry is preserved which is precisely the case for our analysis as the spin wave excitation are \textit{gapless} \cite{Sachdev:2012dq}.} \cite{Hoyos:2013eza},
\begin{eqnarray}
z T^{0}\ _{0}+\delta^{i}_ {j}T^{j}\ _{i}=0
\end{eqnarray}
which for the present case yields,
\begin{eqnarray}
z \epsilon =d P_{\perp}+\Delta
\label{e11}
\end{eqnarray}
where, $ d $ is the number of spatial dimensions. This clearly suggests that the corresponding equation of state is modified due to the presence of anisotropy (spin wave excitations) in the system. Suppose for the time being that we are dealing with an uncharged fluid where the temperature ($ T $) is the only scale in our theory. Therefore, it turns out that the above equation of state (\ref{e11}) could be obtained
from the Euler relation,
\begin{eqnarray}
\epsilon +P_{\perp} = T \mathfrak{s}
\end{eqnarray}
if we demand the following non trivial scaling relations,
\begin{eqnarray}
\epsilon \sim P_{\perp, \parallel} \sim T^{\frac{z+d}{z}+\theta}
\end{eqnarray}
where the exponent, $ \theta \left( = \frac{\Delta}{z P_{\perp}}\right)  $ clearly vanishes near the fixed point of the theory and as a result one recovers the standard scaling relations \cite{Hoyos:2013eza}.

We now focus on the dissipative part of the stress tensor (\ref{E3}).  After a straightforward computation, one could in fact express the dissipative contributions to the stress tensor (\ref{E3}) into following two pieces namely\footnote{Note that, here we have assumed, $ \partial . \mathfrak{v}=0 $ as there do not exist any sources for anisotropy \cite{Gahramanov:2012wz}-\cite{Florkowski:2008ag}.},
\begin{eqnarray}
\mathfrak{T}^{\mu \nu}_{(D)}=\mathfrak{T}^{(0)\mu \nu}_{(D)}+\Delta \mathfrak{T}^{(\Delta)\mu \nu}_{(D)}
\end{eqnarray}
where the first piece\footnote{Here, $ \omega^{\mu}=\frac{1}{2}\varepsilon^{\mu \nu \rho\sigma} \mathfrak{u}_{\nu}\partial_{\rho}\mathfrak{u}_{\sigma}$ is the so called \textit{vorticity} factor \cite{Banerjee:2008th}-\cite{Erdmenger:2008rm}.},
\begin{eqnarray}
\mathfrak{T}^{(0)\mu \nu}_{(D)}&=&\mathfrak{S}^{(0)\mu \nu}+\mathfrak{u}^{\mu}\mathfrak{G}^{(0)\nu}+\mathcal{O}(\partial^{2})\nonumber\\
\mathfrak{S}^{(0)\mu \nu}&=&-\eta \sigma_{(\mathfrak{u})}^{\mu \nu}-\zeta \mathfrak{P}^{\mu \nu}(\partial . \mathfrak{u})\nonumber\\
\mathfrak{G}^{(0)\mu}&=&-\alpha_{1}\mathfrak{u}^{\sigma}\partial_{\sigma}\mathfrak{u}^{\mu}-2\alpha_{2}\left( \mathfrak{E}^{\mu}-T \mathfrak{P}^{\mu \sigma}\partial_{\sigma}\left(\frac{\mu_{c}}{T} \right) \right) -\beta_{\omega}T\omega^{\mu}-\beta_{\mathfrak{B}}T\mathfrak{B}^{\mu}
\label{E10}
\end{eqnarray}
contains the usual contributions to the stress tensor that do not include any effects of anisotropy \cite{Hoyos:2015lra}. On the other hand, the second piece\footnote{The transport associated with the $ \Theta $ term should be considered in a more general sense. When we discuss the parity even sector it should be thought of as the parity even transport while for the parity odd sector we would replace it accordingly by parity odd transports.},
\begin{eqnarray*}
\mathfrak{T}^{(\Delta)\mu \nu}_{(D)}=\mathfrak{S}^{(\Delta)\mu \nu}+\mathfrak{Q}^{(\Delta)\mu \nu}+\mathfrak{u}^{\mu}\mathfrak{G}^{(\Delta)\nu}+\mathcal{O}(\partial^{2})
\end{eqnarray*}
\begin{eqnarray*}
\mathfrak{S}^{(\Delta)\mu \nu} = - \lambda\sigma^{\mu \nu}_{(\mathfrak{v})} -\mathfrak{P}^{\mu \nu}((\mathfrak{v} . \mathfrak{G})-\kappa_{2}\Delta ((\partial. \mathfrak{u})+\Theta)) +\mathfrak{v}^{\mu}\mathfrak{G}^{\nu}+\mathfrak{v}^{\nu}\mathfrak{G}^{\mu} \nonumber\\
-\mathfrak{v}^{\mu}\mathfrak{v}^{\nu} \left( (\mathfrak{v} . \mathfrak{G})-\kappa_{2}\Delta \Theta +\xi_{1}\mathfrak{u}.\partial\left(\frac{\mu_{c}}{T} \right) -\xi_{2}\mathfrak{u}^{\lambda}\mathfrak{u}^{\sigma}
\partial_{\lambda}\mathfrak{v}_{\sigma}+2\kappa_{2}\Delta ((\partial . \mathfrak{u})+\Theta)\right)
\end{eqnarray*}
\begin{eqnarray}
\mathfrak{Q}^{(\Delta)\mu \nu}&= &\mathfrak{u}^{\mu} \mathfrak{v}^{\nu}\left( (\mathfrak{v} . \mathfrak{G})-\kappa_{2}\Delta \Theta +\kappa_{3}\mathfrak{u}.\partial\left( \frac{\mu_{c}}{T}\right)\right) \nonumber\\
\mathfrak{G}^{(\Delta)\mu}&=&\kappa_{1}\mathfrak{v}^{\sigma}\partial_{\sigma}\mathfrak{u}^{\mu}
+\kappa_{2}\mathfrak{v}^{\mu}((\partial . \mathfrak{u})+\Theta)
\label{E11}
\end{eqnarray}
contains all the first order dissipative corrections to the stress tensor (\ref{E3}) that explicitly include the effects of anisotropy. At this stage it is customary to note that in order to arrive at the above relations (\ref{E10})-(\ref{E11}), we have used the following definitions and/or the identifications namely,
\begin{eqnarray}
\mathfrak{S}^{\mu \nu} &=& \mathfrak{S}^{(0)\mu \nu}+\Delta \mathfrak{S}^{(\Delta)\mu \nu}\nonumber\\
\mathfrak{G}^{\mu} &=& \mathfrak{G}^{(0)\mu}+\Delta \mathfrak{G}^{(\Delta)\mu}\nonumber\\
\sigma^{\mu \nu}_{(\mathfrak{x})}&=&\mathfrak{P}^{\mu \sigma}\mathfrak{P}^{\nu \lambda}\left( \partial_{\sigma}\mathfrak{x}_{\lambda}+\partial_{\lambda}\mathfrak{x}_{\sigma}-\delta_{\mathfrak{x},\mathfrak{u}}\frac{2}{3}\mathfrak{P}_{\sigma \lambda}(\partial . \mathfrak{x})\right),~~(\mathfrak{x}=\mathfrak{u},\mathfrak{v})\nonumber\\
\Theta &=& \mathfrak{v}^{\sigma}\left( -\mathfrak{E}_{\sigma}- \mathfrak{B}_{\sigma}+ T \partial_{\sigma}(\mu_{c} / T)- \omega_{\sigma}\right) 
\end{eqnarray}
with, $  \mathfrak{P}^{\mu \nu} = \mathfrak{u}^{\mu}\mathfrak{u}^{\nu}+\mathfrak{g}^{\mu \nu}$ as the projection operator. At this stage it is also noteworthy to mention that the entity, $ \mathfrak{G}^{\mu} $ contains all the first order dissipative corrections in the basic hydrodynamic variables namely, the fluid velocity ($ \mathfrak{u}^{\mu} $), chemical potential ($ \mu $) and the temperature ($ T $). Moreover, here the entity, $\mathfrak{u}_{\mu}\mathfrak{G}^{\mu}=0 $ in the Landau frame \cite{Hoyos:2015lra}.

Finally, the $ U(1) $ charged current could be formally expressed as\footnote{In principle one could have added a term proportional to $ \mathfrak{v}^{\mu} $ at the zeroth order level in the derivative expansion. However, as far as the present analysis is concerned we switch off all the components of the \textit{electric} current along the direction of the anisotropic vector field \cite{Gahramanov:2012wz}-\cite{Florkowski:2008ag}.},
\begin{eqnarray}
\mathfrak{j}^{a \mu}&=&\varrho^{a}\mathfrak{u}^{\mu}+\mathfrak{D}^{a \mu}+\mathcal{O}(\partial^{2})
\end{eqnarray}
where, the function $\mathfrak{D}^{ a\mu}  $ contains all possible dissipative corrections in the first order derivative expansion namely\footnote{One could formally express, $\mathfrak{D}^{a \mu} \equiv \mathfrak{D}^{(0)a\mu}+\Delta \mathfrak{D}^{(\Delta)a\mu} $.},
\begin{eqnarray}
\mathfrak{D}^{a \mu}&=&2\alpha^{a}_{3}\mathfrak{u}^{\sigma}\partial_{\sigma}\mathfrak{u}^{\mu}+(\sigma_{D})^{a}
\ _{b}\left( \mathfrak{E}^{b\mu}-T \mathfrak{P}^{\mu \sigma}\partial_{\sigma}\left(\frac{\mu^{b}_{c}}{T} \right) \right)+\varsigma^{a}_{\omega}\omega^{\mu}+(\varsigma_{\mathfrak{B}})^{a}\ _{b}\mathfrak{B}^{b\mu}+\Delta\mathfrak{D}^{(\Delta)a\mu}\nonumber\\
\mathfrak{D}^{(\Delta)a\mu}&=& \kappa^{a}_{1}\mathfrak{v}^{\sigma}
\partial_{\sigma}\mathfrak{u}^{\mu}-\mathfrak{v}^{\mu}\left( (\mathfrak{v} . \mathfrak{G}^{a})-\Delta (\kappa_{2})^{a}\ _{b} \Theta^{b}+\gamma^{a}_{1}\mathfrak{v}^{\lambda}\mathfrak{u}^{\sigma}\partial_{\sigma}
\mathfrak{u}_{\lambda}+(\gamma_{2})^{a}\ _{b}\mathfrak{u}.\partial \left( \frac{\mu^{b}_{c}}{T}\right)\right)   
\label{E14}
\end{eqnarray}
such that in the Landau frame, $ \mathfrak{u}_{\mu}\mathfrak{j}^{a \mu}=-\varrho^{a} $ .

%%%%%%%%%%%%%%%%%%%%%%%%%%%%%%%%%%%%
\section{Entropy current}
The purpose of this Section is to construct a systematic hydrodynamic description away from the quantum critical region with Lifshitz isometry group. In order to do that, in this paper we illustrate the simplest case with, $ N=1 $.

%%%%%%%%%%%%%%%%%%%%%%%%%%%%%%%%%%%%
\subsection{Thermodynamic identities}
As a first step of our analysis, we first construct some basic thermodynamic identities which will be required in the subsequent analysis. Considering the following identity\footnote{In order to avoid confusion, we denote chemical potential as $ \mu_{c} $.},
\begin{eqnarray}
\mathbb{I}=\mathfrak{u}_{\nu}\partial_{\mu}\mathfrak{T}^{\mu \nu}+\mu_{c} \partial_{\mu}\mathfrak{j}^{\mu}
\label{E13}
\end{eqnarray}
and using (\ref{E1}), it is indeed quite trivial to note that, $ \mathbb{I} \sim \mathcal{O}(\partial^{2}) $. Which therefore clearly suggests that at the zeroth order level in the derivative expansion of $ \mathfrak{T}^{\mu \nu} $ and $ \mathfrak{j}^{\mu} $ one must have, $ \mathbb{I}=0 $ which naturally yields,
\begin{eqnarray}
-\mathfrak{u}^{\mu}\partial_{\mu}\epsilon +\frac{(\epsilon + P_{\perp})}{\mathfrak{s}}\mathfrak{u}^{\mu}\partial_{\mu}\mathfrak{s}
-\Delta\mathfrak{u}_{\nu}\mathfrak{v}^{\mu}\partial_{\mu}\mathfrak{v}^{\nu}+\mu_{c} \partial_{\mu}\varrho \mathfrak{u}^{\mu}-\frac{\mu_{c} \varrho}{\mathfrak{s}}\mathfrak{u}^{\mu}\partial_{\mu}\mathfrak{s}=0
\label{E14}
\end{eqnarray}
where, $ \mathfrak{s} $ is the so called entropy density such that, $ \partial_{\mu}(\mathfrak{s}\mathfrak{u}^{\mu})=0 $. 

We would now like to interpret the entity $ \mathfrak{v}^{\mu}\partial_{\mu}\mathfrak{v}^{\nu} $ above in (\ref{E14}). To do that, let us first note that in the so called \textit{boosted} frame one could formally express the four velocities as,
\begin{eqnarray}
\mathfrak{u}^{\mu}&=&\left( \frac{1}{\sqrt{1-\beta^{2}}},\frac{\beta^{i}}{\sqrt{1-\beta^{2}}}\right),~~(i=x,y,z)\nonumber\\
\mathfrak{v}^{\mu}&=&\left( \frac{\beta_{z}}{\sqrt{1-\beta^{2}-\beta_{z}^{2}}},0,0,\frac{\sqrt{1-\beta^{2}}}{\sqrt{1-\beta^{2}-\beta_{z}^{2}}}\right)
\label{E15}
\end{eqnarray}
which clearly satisfy the above constraint in (\ref{E6}). Here, $ \beta^{i} (=v^{i}/c)$ are the arbitrary boosts along three different spatial directions. Using (\ref{E15}), it is indeed quite straight forward to show,
\begin{eqnarray}
\mathfrak{v}^{\mu}\partial_{\mu}\mathfrak{v}^{\nu} =\partial^{\nu}\log \vartheta (\beta)
\label{e18}
\end{eqnarray}
where the function $ \vartheta(\beta) $ could be formally expressed as,
\begin{eqnarray}
\vartheta (\beta)=\frac{\beta_{z}}{{\sqrt{1-\beta^{2}-\beta_{z}^{2}}}} \delta_{\nu , 0}+\frac{\sqrt{1-\beta^{2}}}{{\sqrt{1-\beta^{2}-\beta_{z}^{2}}}} \delta_{\nu , z}.
\label{E17}
\end{eqnarray}

Using (\ref{E17}), one could now formally express (\ref{E14}) as, 
\begin{eqnarray}
-\mathfrak{u}^{\mu}\partial_{\mu}\epsilon +\frac{(\epsilon + P_{\perp})}{\mathfrak{s}}\mathfrak{u}^{\mu}\partial_{\mu}\mathfrak{s}
-\frac{\Delta}{\vartheta}\mathfrak{u}^{\nu}\partial_{\nu}\vartheta+\mu_{c} (\partial_{\mu}\varrho )\mathfrak{u}^{\mu}-\frac{\mu_{c} \varrho}{\mathfrak{s}}\mathfrak{u}^{\mu}\partial_{\mu}\mathfrak{s}=0.
\label{E18}
\end{eqnarray}

Following the arguments of \cite{Gahramanov:2012wz}-\cite{Florkowski:2008ag}, next we introduce the so called \textit{generalized} energy density function,
\begin{eqnarray}
\epsilon = \epsilon (\varrho , \mathfrak{s}, \vartheta)
\end{eqnarray}
which naturally yields,
\begin{eqnarray}
d \epsilon &=&\left( \frac{\partial \epsilon}{\partial \mathfrak{s}}\right)_{\varrho , \vartheta} d\mathfrak{s}+\left( \frac{\partial \epsilon}{\partial \varrho}\right)_{\mathfrak{s},\vartheta} d\varrho +\left( \frac{\partial \epsilon}{\partial \vartheta}\right)_{\mathfrak{s},\varrho}  d\vartheta \nonumber\\
&=& T d \mathfrak{s}+\mu_{c} d\varrho - \frac{\Delta}{\vartheta} d\vartheta.
\label{E20}
\end{eqnarray}
The above equation (\ref{E20}) could be thought of as the modified first law of thermodynamics in the presence of anisotropies. The first two terms are precisely the contributions arising from the entropy density as well as the charge density associated with the system. Whereas, on the other hand, the last term is precisely the anisotropic contribution to the energy density of the system. 

Finally, using (\ref{E18}) and (\ref{E20}), it is quite useful to obtain the following set of identities,
\begin{eqnarray}
\epsilon + P_{\perp}&=&T \mathfrak{s}+\mu_{c} \varrho\nonumber\\
d P_{\perp}&=& \mathfrak{s} dT+\varrho d\mu_{c} +\frac{\Delta}{\vartheta} d\vartheta
\label{E21}
\end{eqnarray}
which we use in the subsequent discussions. 

Before we conclude this Section, it is now customary to derive the so called equation for the entropy current \cite{Son:2009tf} in the presence of the pressure anisotropy. Using (\ref{E1}), it is indeed quite trivial to note down the following conservation equation, 
\begin{eqnarray}
\mathfrak{u}_{\nu}\partial_{\mu}\mathfrak{T}^{\mu \nu}+\mu_{c} \partial_{\mu}\mathfrak{j}^{\mu}=-\mathfrak{E}^{\mu}\mathfrak{j}_{\mu}+\mu_{c} \mathfrak{c}\mathfrak{E}^{\mu}\mathfrak{B}_{\mu}
\end{eqnarray}
which by virtue of the arguments as mentioned above in (\ref{E13}) gives rise to non trivial contributions at the quadratic order in the derivative expansion namely,
\begin{eqnarray}
\partial_{\mu}\mathbb{J}^{\mu}_{\mathfrak{s}}=-\frac{1}{T}\partial_{\mu}\mathfrak{u}_{\nu}\mathfrak{S}^{(0)\mu \nu}-\mathfrak{D}^{(0)\mu}\left(\mathfrak{P}_{\mu \nu}\partial^{\nu}\left(\frac{\mu_{c}}{T} \right)- \frac{\mathfrak{E}_{\mu}}{T} \right) -\frac{\mu_{c}}{T}\mathfrak{c}\mathfrak{E}.\mathfrak{B}-\frac{1}{T}\mathfrak{G}^{(0)\nu}\mathfrak{u}^{\mu}\partial_{\mu}\mathfrak{u}_{\nu}
+\Delta\mathbb{M}^{(\Delta)}
\label{E24}
\end{eqnarray}
where, 
\begin{eqnarray}
\mathbb{J}^{\mu}_{\mathfrak{s}}=\mathfrak{s}\mathfrak{u}^{\mu}-\frac{\mu_{c}}{T}\mathfrak{D}^{\mu}=\mathbb{J}^{(0)\mu}_{\mathfrak{s}}+\Delta \mathbb{J}^{(\Delta)\mu}_{\mathfrak{s}}
\label{E25}
\end{eqnarray}
is the so called entropy current of the system. Clearly, from (\ref{E25}), it is quite evident that in the presence of pressure anisotropies, the so called entropy current of the system receives non trivial corrections which is naturally reflected in its divergence equation (\ref{E24}). The first part on the R.H.S. of (\ref{E24}) is the usual contribution to the entropy current in the absence of pressure anisotropy \cite{Hoyos:2015lra}. The entity $ \mathbb{M}^{(\Delta)} $, on the other hand, summaries all the anisotropic contributions to the divergence equation namely,
\begin{eqnarray}
\mathbb{M}^{(\Delta)}=-\frac{1}{T}\partial_{\mu}\mathfrak{u}_{\nu}\mathfrak{S}^{(\Delta)\mu \nu}-\mathfrak{D}^{(\Delta)\mu}\left(\mathfrak{P}_{\mu \nu}\partial^{\nu}\left(\frac{\mu_{c}}{T} \right)- \frac{\mathfrak{E}_{\mu}}{T} \right)-\frac{1}{T}\mathfrak{G}^{(\Delta)\nu}\mathfrak{u}^{\mu}\partial_{\mu}\mathfrak{u}_{\nu}
+\frac{1}{T}\mathfrak{u}_{\nu}\partial_{\mu}\mathfrak{Q}^{(\Delta)\mu \nu}.
\end{eqnarray} 
%%%%%%%%%%%%%%%%%%%%%%%%%%%%%
\subsection{A note on the parity even sector}
Before we proceed further, at this stage it is noteworthy to mention that the parity breaking terms would definitely modify the entropy current as expected from the earlier analysis \cite{Son:2009tf}. However, before moving to the parity \textit{odd} sector, it is customary to check the parity even sector in the presence of pressure anisotropy\footnote{In the subsequent discussion of this subsection we consider only the parity even part of (\ref{E24}). In principle, one could do this since the two sectors do not get mixed up even in the presence of anisotropy \cite{Hoyos:2015lra}. A careful investigation would reveal the fact that this is the unique feature of first order dissipative hydrodynamics which essentially truncates constitutive relations upto leading order in $ \mathfrak{G}^{\mu} $.}. As the reader might have noticed so far that due to the presence of the pressure anisotropy in the system, even in the parity even sector of the stress tensor (\ref{E11}) (as well as the charge current (\ref{E14})) we encounter a number of additional transports which need to be fixed first. We expect to fix these coefficients purely from the condition of the positivity of the entropy current namely, $ \partial_{\mu}\mathbb{J}^{\mu}_{\mathfrak{s}} \geq 0 $. 

We would first consider the parity \textit{even} sector in (\ref{E24}) namely,
\begin{eqnarray}
\partial_{\mu}\mathbb{J}^{(P)\mu}_{\mathfrak{s}}=-\frac{1}{T}\partial_{\mu}\mathfrak{u}_{\nu}\mathfrak{S}^{(0)\mu \nu}-\mathfrak{D}^{(0,P)\mu}\left(\mathfrak{P}_{\mu \nu}\partial^{\nu}\left(\frac{\mu_{c}}{T} \right)- \frac{\mathfrak{E}_{\mu}}{T} \right) -\frac{1}{T}\mathfrak{G}^{(0,P)\nu}\mathfrak{u}^{\mu}\partial_{\mu}\mathfrak{u}_{\nu}
+\Delta\mathbb{M}^{(\Delta ,P)}
\label{E28}
\end{eqnarray}
where each of the individual entities could be formally expressed as,
\begin{eqnarray*}
\mathbb{M}^{(\Delta ,P)}=-\frac{1}{T}\partial_{\mu}\mathfrak{u}_{\nu}\mathfrak{S}^{(\Delta , P)\mu \nu}-\mathfrak{D}^{(\Delta , P)\mu}\left(\mathfrak{P}_{\mu \nu}\partial^{\nu}\left(\frac{\mu_{c}}{T} \right)- \frac{\mathfrak{E}_{\mu}}{T} \right)-\frac{1}{T}\mathfrak{G}^{(\Delta , P)\nu}\mathfrak{u}^{\mu}\partial_{\mu}\mathfrak{u}_{\nu}
+\frac{1}{T}\mathfrak{u}_{\nu}\partial_{\mu}\mathfrak{Q}^{(\Delta ,P)\mu \nu}
\end{eqnarray*}
\begin{eqnarray*}
\mathfrak{S}^{(\Delta , P)\mu \nu}=- \lambda\sigma^{\mu \nu}_{(\mathfrak{v})} -\mathfrak{P}^{\mu \nu}((\mathfrak{v} . \mathfrak{G}^{(P)})-\kappa_{2}\Delta ((\partial. \mathfrak{u})+\Theta^{(P)})) +\mathfrak{v}^{\mu}\mathfrak{G}^{(P)\nu}+\mathfrak{v}^{\nu}\mathfrak{G}^{(P)\mu} \nonumber\\
-\mathfrak{v}^{\mu}\mathfrak{v}^{\nu} \left( (\mathfrak{v} . \mathfrak{G}^{(P)})-\kappa_{2}\Delta \Theta^{(P)} +\xi_{1}\mathfrak{u}.\partial\left(\frac{\mu_{c}}{T} \right) -\xi_{2}\mathfrak{u}^{\lambda}\mathfrak{u}^{\sigma}
\partial_{\lambda}\mathfrak{v}_{\sigma}+2\kappa_{2}\Delta ((\partial . \mathfrak{u})+\Theta^{(P)})\right)
\end{eqnarray*}
\begin{eqnarray*}
\mathfrak{D}^{(0,P)\mu}=2\alpha_{3}\mathfrak{u}^{\sigma}\partial_{\sigma}\mathfrak{u}^{\mu}
+\sigma_{D}\left( \mathfrak{E}^{\mu}-T \mathfrak{P}^{\mu \sigma}\partial_{\sigma}\left(\frac{\mu_{c}}{T} \right) \right)
\end{eqnarray*}
\begin{eqnarray*}
 \mathfrak{D}^{(\Delta , P)\mu}=\kappa_{1}\mathfrak{v}^{\sigma}
\partial_{\sigma}\mathfrak{u}^{\mu}-\mathfrak{v}^{\mu}\left( (\mathfrak{v} . \mathfrak{G}^{(P)})-\kappa_{2}\Delta \Theta^{(P)}+\gamma_{1}\mathfrak{v}^{\lambda}\mathfrak{u}^{\sigma}\partial_{\sigma}
\mathfrak{u}_{\lambda}+\gamma_{2} \mathfrak{u}.\partial \left( \frac{\mu^{b}_{c}}{T}\right)\right)
\end{eqnarray*}
\begin{eqnarray*}
\mathfrak{Q}^{(\Delta ,P)\mu \nu}= \mathfrak{u}^{\mu} \mathfrak{v}^{\nu}\left( (\mathfrak{v} . \mathfrak{G}^{(P)})-\kappa_{2}\Delta \Theta^{(P)} +\kappa_{3}\mathfrak{u}.\partial\left( \frac{\mu_{c}}{T}\right)\right)
\end{eqnarray*}
\begin{eqnarray}
\mathfrak{G}^{(P)\mu}&=&\mathfrak{G}^{(0,P)\mu}+\Delta \mathfrak{G}^{(\Delta ,P)\mu}\nonumber\\
\mathfrak{G}^{(0,P)\mu}&=& -\alpha_{1}\mathfrak{u}^{\sigma}\partial_{\sigma}\mathfrak{u}^{\mu}-2\alpha_{2}\left( \mathfrak{E}^{\mu}-T \mathfrak{P}^{\mu \sigma}\partial_{\sigma}\left(\frac{\mu_{c}}{T} \right) \right)\nonumber\\
\mathfrak{G}^{(\Delta ,P)\mu} &=&\kappa_{1}\mathfrak{v}^{\sigma}\partial_{\sigma}\mathfrak{u}^{\mu}
+\kappa_{2}\mathfrak{v}^{\mu}((\partial . \mathfrak{u})+\Theta^{(P)})\nonumber\\
\Theta^{(P)}&=& \mathfrak{v}^{\sigma}\left( -\mathfrak{E}_{\sigma}+ T \partial_{\sigma}(\mu_{c} / T)\right).
\end{eqnarray}

At this stage, it is also noteworthy to mention that some of the transports appearing in the isotropic part of the parity even sector also appears in its anisotropic counterpart. As a consequence of this, all the previously determined constraints \cite{Hoyos:2015lra} on the transport coefficients of the (isotropic) parity even sector (in the constitutive relations) might not hold good and one therefore needs to redo the calculations. In the following we enumerate each of the entities in (\ref{E28}) separately. A straightforward computation yields the following:\\ \\
%%%%%%%%%%%%%%%%%%%%%%%%%%%%%%%%%%
\textbf{(I)}  
\begin{eqnarray}
-\frac{1}{T}\partial_{\mu}\mathfrak{u}_{\nu}\mathfrak{S}^{(0)\mu \nu} =\frac{\eta}{T}\sigma^{\mu \nu}\sigma_{\mu \nu}+\frac{\zeta}{T}(\partial . \mathfrak{u})^{2}.
\label{E30}
\end{eqnarray}
\textbf{(II)}  
\begin{eqnarray*}
-\mathfrak{D}^{(0,P)\mu}\left(\mathfrak{P}_{\mu \nu}\partial^{\nu}\left(\frac{\mu_{c}}{T} \right)- \frac{\mathfrak{E}_{\mu}}{T} \right)  =\frac{\sigma_{D}}{T}\left(\mathfrak{E}_{\mu}-T\mathfrak{P}_{\mu \nu}\partial^{\nu}\left(\frac{\mu_{c}}{T} \right)\right)^{2} +\frac{\alpha_{3}}{T}\mathfrak{P}^{\mu \nu}\mathrm{K}_{\mu \nu}\Theta^{(P)}
\end{eqnarray*}
where,
\begin{eqnarray}
\mathrm{K}_{\alpha \beta}=\partial_{\alpha}\mathfrak{v}_{\beta}+\partial_{\beta}\mathfrak{v}_{\alpha}.
\end{eqnarray}
\textbf{(III)}  
\begin{eqnarray}
-\frac{1}{T}\mathfrak{G}^{(0,P)\nu}\mathfrak{u}^{\mu}\partial_{\mu}\mathfrak{u}_{\nu}
=\frac{\alpha_{1}}{T}(\mathfrak{u}^{\sigma}\partial_{\sigma}\mathfrak{u}^{\mu})^{2}+\frac{\alpha_{2}}{T}\mathfrak{P}^{\mu\nu}\mathrm{K}_{\mu \nu}\Theta^{(P)}.
\end{eqnarray}
\textbf{(IV)} 
\begin{eqnarray*}
-\frac{1}{T}\partial_{\mu}\mathfrak{u}_{\nu}\mathfrak{G}^{(\Delta , P)\mu \nu}=\frac{\lambda \mathrm{K}^{\mu \nu}}{T}(\partial_{\sigma}\mathfrak{u}_{\nu}\mathfrak{u}_{\mu}u^{\sigma}+\frac{\mathrm{Q}_{\mu \nu}}{2})
+\frac{\alpha_{1}}{T}(\partial . \mathfrak{u})\mathfrak{P}^{\mu \nu}\mathrm{K}_{\mu \nu}+\frac{\xi_{1}}{2T}(\mathfrak{v}^{\mu}\mathfrak{v}^{\nu}\mathrm{Q}_{\mu \nu})\left(\mathfrak{u}.\partial \left( \frac{\mu_{c}}{T}\right)  \right) \nonumber\\
+\frac{\alpha_{1}}{T}\mathfrak{v}^{\mu}\mathfrak{u}^{\sigma}\mathrm{Q}_{\mu \nu}\partial_{\sigma}\mathfrak{u}^{\nu}+\frac{2\alpha_{2}}{T}(\partial . \mathfrak{u})\Theta^{(P)}+\frac{\Delta (\kappa_{1}-\kappa_{2})}{2T}(\partial . \mathfrak{u})\mathfrak{v}^{\mu}\mathfrak{v}^{\nu}\mathrm{Q}_{\mu \nu}-\frac{\alpha_{2}}{T}\mathfrak{v}^{\mu}\mathfrak{v}^{\nu}\mathrm{Q}_{\mu \nu} \Theta^{(P)}\nonumber\\
-\frac{\Delta \kappa_{1}}{T}\mathfrak{v}^{\mu}\mathfrak{v}^{\sigma}\mathrm{Q}_{\mu \nu}\partial_{\sigma}\mathfrak{u}^{\nu}+\frac{(\alpha_{1}-\xi_{2})}{4T}\mathfrak{P}^{\sigma \lambda}\mathrm{K}_{\sigma \lambda}\mathfrak{v}^{\mu}\mathfrak{v}^{\nu}\mathrm{Q}_{\mu \nu}+\frac{\Delta \kappa_{1}}{4T}(\mathfrak{v}^{\mu}\mathfrak{v}^{\nu}\mathrm{Q}_{\mu \nu})^{2}
\end{eqnarray*}
where,
\begin{eqnarray}
\mathrm{Q}_{\alpha \beta}=\partial_{\alpha}\mathfrak{u}_{\beta}
+\partial_{\beta}\mathfrak{u}_{\alpha}.
\end{eqnarray}
\textbf{(V)}
\begin{eqnarray}
-\mathfrak{D}^{(\Delta , P)\mu}\left(\mathfrak{P}_{\mu \nu}\partial^{\nu}\left(\frac{\mu_{c}}{T} \right)- \frac{\mathfrak{E}_{\mu}}{T} \right)=-\frac{\kappa_{1}(1-\Delta)}{2T}\mathfrak{v}^{\mu}\mathfrak{v}^{\nu}\mathrm{Q}_{\mu \nu}\Theta^{(P)}+\frac{2\alpha_{2}}{T}\Theta^{2(P)}\nonumber\\+\frac{(\alpha_{1}-\gamma_{1})}{2T}\mathfrak{P}^{\mu \nu}\mathrm{K}_{\mu \nu}\Theta^{(P)}
+\frac{\Delta \kappa_{2}}{T}(\partial . \mathfrak{u})\Theta^{(P)}+\frac{\gamma_{2}}{T}\Theta^{(P)}\mathfrak{u}. \partial\left(\frac{\mu_{c}}{T} \right).
\end{eqnarray}
\textbf{(VI)}
\begin{eqnarray}
-\frac{1}{T}\mathfrak{G}^{(\Delta , P)\nu}\mathfrak{u}^{\mu}\partial_{\mu}\mathfrak{u}_{\nu}=-\frac{\kappa_{1}}{T}\mathfrak{v}^{\mu}\mathfrak{u}^{\sigma}\mathrm{Q}_{\mu \nu}\partial_{\sigma}\mathfrak{u}^{\nu}-\frac{\kappa_{1}}{T}\mathfrak{u}^{\mu}\mathfrak{u}^{\sigma}\mathrm{K}_{\mu \nu}\partial_{\sigma}
\mathfrak{u}^{\nu}-\frac{(\kappa_{1}-\kappa_{2})}{2T}(\partial . \mathfrak{u})\mathfrak{P}^{\mu \nu}\mathrm{K}_{\mu \nu}\nonumber\\
+\frac{\kappa_{1}}{T}\mathfrak{u}^{\mu}\partial_{\mu}\mathfrak{v}_{\nu}\partial_{\sigma}
(\mathfrak{u}^{\sigma}\mathfrak{u}^{\nu})
+\frac{\kappa_{2}}{2T}\mathfrak{P}^{\mu \nu}\mathrm{K}_{\mu \nu}\Theta^{(P)}.
\end{eqnarray}
\textbf{(VII)}
\begin{eqnarray}
\frac{1}{T}\mathfrak{u}_{\nu}\partial_{\mu}\mathfrak{Q}^{(\Delta ,P)\mu \nu}=\frac{\alpha_{1}}{4T}\left( \mathfrak{P}^{\mu \nu}\mathrm{K}_{\mu \nu}\right)^{2}
+\frac{\alpha_{2}}{T}\mathfrak{P}^{\mu \nu}\mathrm{K}_{\mu \nu}\Theta^{(P)}+\frac{\Delta \kappa_{1}}{4T}\mathfrak{P}^{\lambda \sigma}\mathrm{K}_{\lambda \sigma}\mathfrak{v}^{\mu}\mathfrak{v}^{\nu}\mathrm{Q}_{\mu \nu}
\nonumber\\+\frac{\Delta \kappa_{2}}{2T}(\partial . \mathfrak{u})\mathfrak{P}^{\mu \nu}\mathrm{K}_{\mu \nu}+\frac{\kappa_{3}}{2T}\left( \mathfrak{P}^{\mu \nu}\mathrm{K}_{\mu \nu}\right)\left(\mathfrak{u} . \partial\left(\frac{\mu_{c}}{T} \right)  \right) . 
\label{E36}
\end{eqnarray}
%%%%%%%%%%%%%%%%%%%%%%%%%%%%%%%%
$ \bullet $ \textbf{Observations:} In order for the entropy current (\ref{E28}) to be positive definite, from the above computations (\ref{E30})-(\ref{E36}) we observe the following\footnote{In the following discussion we assume, $ \Delta>0 $.}\\
(1) Three of the following transports need to be positive definite namely,
\begin{eqnarray}
\eta \geq 0,~~\zeta \geq 0,~~\sigma_{D}\geq 0.
\end{eqnarray}
(2) The rest of the transports satisfy nontrivial algebraic identities part of which could be simplified further in order to obtain,
\begin{eqnarray}
\lambda &= & \alpha_{2}=\alpha_{1},~~\kappa_{2}=\kappa_{1},~~\xi_{2}=\kappa_{1}(2\Delta -1)-\alpha_{1},\nonumber\\
\Delta \gamma_{1}-2\alpha_{3}&=&(3\Delta +2)\alpha_{1}+\Delta \kappa_{1} \geq 0,~~\alpha_{1},\alpha_{2}\geq 0,~~\kappa_{1}\geq 0.
\label{E38}
\end{eqnarray}
(3) The rest of the constraints could be formally expressed as,
\begin{eqnarray}
\lambda \mathrm{K}^{\mu \nu}\mathrm{Q}_{\mu \nu}+2\kappa_{1}\left( \mathfrak{P}^{\mu \nu}\mathrm{K}_{\mu \nu}(\partial . \mathfrak{u})+(\mathfrak{u}^{\mu}\partial_{\mu}\mathfrak{v}_{\nu})(\mathfrak{u}^{\sigma}\partial_{\sigma}\mathfrak{u}^{\nu})-\Delta \mathfrak{v}^{\mu}(\mathfrak{v}^{\sigma}\partial_{\sigma}\mathfrak{u}^{\nu})\mathrm{Q}_{\mu \nu}\right)&=&0\nonumber\\ 
\xi_{1}(\mathfrak{v}^{\mu}\mathfrak{v}^{\nu}\mathrm{Q}_{\mu \nu})+2\gamma_{2}\Theta^{(P)}+\kappa_{3}(\mathfrak{P}^{\mu \nu}\mathrm{K}_{\mu \nu})&=&0.
\label{E39}
\end{eqnarray}
In other words, once we move away from the quantum criticality (by breaking symmetries associated to it) a number of additional transports appear which are found to obey non trivial constraints among themselves. However, all of these new transports 
cannot be determined uniquely as we have lesser number of constraints available. It turns out that once we determine a few of them, the rest is trivially fixed due to constraints (\ref{E38})-(\ref{E39}) present in the system.

%%%%%%%%%%%%%%%%%%%%%%%%%%%%%%%%%%%%
\subsection{Some useful relations}
We now focus solely on the parity \textit{odd} sector of (\ref{E24}). Before we proceed further, it is now customary to derive certain useful identities that would be required in the subsequent analysis of the entropy current.\\ \\
%%%%%%%%%%%%%%%%%%%%%%%%%%%%%%%
$ \bullet $~~\underline{\textbf{Derivation of $ \partial_{\mu}\omega^{\mu} $:}} We start our analysis by noting down the following identity\cite{Hoyos:2015lra},
\begin{eqnarray}
\mathfrak{u}^{\mu}\mathfrak{u}^{\nu}\partial_{\mu}\omega_{\nu}=-\frac{1}{2}\partial_{\mu}\omega^{\mu}.
\end{eqnarray}

Next, using conservation equation (\ref{E1}) we find,
\begin{eqnarray}
\omega_{\nu}\partial_{\mu}\mathfrak{T}^{\mu \nu}=\varrho \omega_{\mu}\mathfrak{E}^{\mu}+\mathcal{O}(\partial^{3}).
\label{E41}
\end{eqnarray}

Finally, using (\ref{E5}), (\ref{e18}) and (\ref{E41}) we obtain,
\begin{eqnarray}
\partial_{\mu}\omega^{\mu}=-\frac{2}{(\epsilon +P_{\perp})}\omega^{\mu}\mathbb{W}_{\mu}+\mathcal{O}(\partial^{3})
\end{eqnarray}
where, the vector $ \mathbb{W}^{\mu} $ could be formally expressed as,
\begin{eqnarray}
\mathbb{W}^{\mu}=g^{\mu \nu}\left( \partial_{\nu}P_{\perp}-\Delta \partial_{\nu} \log\vartheta \right)-\varrho \mathfrak{E}^{\mu}-\mathfrak{v}^{\mu}\mathfrak{v}^{\nu}\partial_{\nu}\Delta .
\end{eqnarray}
%%%%%%%%%%%%%%%%%%%%%%%%%%%%%%%%%%
$ \bullet $~~\underline{\textbf{Derivation of $ \partial_{\mu}\mathfrak{B}^{\mu} $:}} 
We start our analysis with the following identity \cite{Hoyos:2015lra},
\begin{eqnarray}
\partial_{\mu}\mathfrak{B}^{\mu}=\mathfrak{u}^{\nu}\partial_{\nu}\mathfrak{u}_{\mu}\mathfrak{B}^{\mu}
-2\omega_{\mu}\mathfrak{E}^{\mu}=-\mathfrak{u}^{\mu}\mathfrak{u}^{\nu}\partial_{\mu}
\mathfrak{B}_{\nu}-2\omega_{\mu}\mathfrak{E}^{\mu}
\label{E44}
\end{eqnarray}

Following the same steps as above we note that,
\begin{eqnarray}
\mathfrak{B}_{\nu}\partial_{\mu}\mathfrak{T}^{\mu \nu}=\varrho \mathfrak{B}_{\mu}\mathfrak{E}^{\mu} + \mathcal{O}(\partial^{3}).
\end{eqnarray}

Finally, using (\ref{E5}), (\ref{e18}) and (\ref{E44}) we find,
\begin{eqnarray}
\partial_{\mu}\mathfrak{B}^{\mu}= -2\omega_{\mu}\mathfrak{E}^{\mu}-\frac{1}{(\epsilon +P_{\perp})}\mathfrak{B}^{\mu}\mathbb{W}_{\mu}+\mathcal{O}(\partial^{3}).
\end{eqnarray}

%%%%%%%%%%%%%%%%%%%%%%%%%%%%%%%%%%%%
\subsection{Parity odd sector}
Following the prescription of \cite{Son:2009tf}, we modify our entropy current as,
\begin{eqnarray}
\mathbb{J}^{\mu}_{\mathfrak{s}} \rightarrow \mathbb{J}^{\mu}_{\mathfrak{s}} +\mathbb{D}_{\omega} \omega^{\mu}+ \mathbb{D}_{\mathfrak{B}}\mathfrak{B}^{\mu}=\mathbb{J}^{\mu}_{\mathfrak{s}} +\tilde{\mathbb{J}}^{\mu}_{\mathfrak{s}}
\end{eqnarray}
which yields the divergence equation as\footnote{At this stage it is noteworthy to mention that here by $ \slashed{P} $ we explicitly mean all the parity \textit{odd} terms of the entropy current (\ref{E24}).},
\begin{eqnarray}
\partial_{\mu}\mathbb{J}^{(\slashed{P})\mu}_{\mathfrak{s}}=-\mathfrak{D}^{(0,\slashed{P})\mu}\left(\mathfrak{P}_{\mu \nu}\partial^{\nu}\left(\frac{\mu_{c}}{T} \right)- \frac{\mathfrak{E}_{\mu}}{T} \right) -\frac{\mu_{c}}{T}\mathfrak{c}\mathfrak{E}_{\mu}\mathfrak{B}^{\mu}-\frac{1}{T}\mathfrak{G}^{(0,\slashed{P})\nu}\mathfrak{u}^{\mu}\partial_{\mu}\mathfrak{u}_{\nu}
+\Delta\mathbb{M}^{(\Delta ,\slashed{P})}+\partial_{\mu}\tilde{\mathbb{J}}^{\mu}_{\mathfrak{s}}
\label{E48}
\end{eqnarray}
where, the function $ \mathbb{M}^{(\Delta ,\slashed{P})} $ could be formally expressed as,
\begin{eqnarray}
\mathbb{M}^{(\Delta ,\slashed{P})}=-\frac{1}{T}\partial_{\mu}\mathfrak{u}_{\nu}\mathfrak{S}^{(\Delta , \slashed{P})\mu \nu}-\mathfrak{D}^{(\Delta , \slashed{P})\mu}\left(\mathfrak{P}_{\mu \nu}\partial^{\nu}\left(\frac{\mu_{c}}{T} \right)- \frac{\mathfrak{E}_{\mu}}{T} \right)-\frac{1}{T}\mathfrak{G}^{(\Delta , \slashed{P})\nu}\mathfrak{u}^{\mu}\partial_{\mu}\mathfrak{u}_{\nu}\nonumber\\
+\frac{1}{T}\mathfrak{u}_{\nu}\partial_{\mu}\mathfrak{Q}^{(\Delta ,\slashed{P})\mu \nu}.
\end{eqnarray}

In the following, we enumerate each of the individual terms one by one.\\ \\
%%%%%%%%%%%%%%%%%%%%%%%%%%%%%
\textbf{(I)}  
\begin{eqnarray}
-\mathfrak{D}^{(0,\slashed{P})\mu}\left(\mathfrak{P}_{\mu \nu}\partial^{\nu}\left(\frac{\mu_{c}}{T} \right)- \frac{\mathfrak{E}_{\mu}}{T} \right)  =\frac{\varsigma_{\omega}}{T}\omega^{\mu}\mathfrak{E}_{\mu}-\varsigma_{\omega}\omega^{\mu}\partial_{\mu}\left( \frac{\mu_{c}}{T}\right)+\frac{\varsigma_{\mathfrak{B}}}{T}\mathfrak{E}_{\mu}\mathfrak{B}^{\mu}-\varsigma_{\mathfrak{B}}\mathfrak{B}^{\mu}\partial_{\mu}\left( \frac{\mu_{c}}{T}\right).
\end{eqnarray}
\textbf{(II)}  
\begin{eqnarray}
-\frac{1}{T}\mathfrak{G}^{(0,\slashed{P})\nu}\mathfrak{u}^{\mu}\partial_{\mu}\mathfrak{u}_{\nu} =\frac{(\beta_{\omega}\omega^{\mu}+\beta_{\mathfrak{B}}\mathfrak{B}^{\mu})}{(\epsilon +P_{\perp})}\left( \varrho \mathfrak{E}_{\mu}-\partial_{\mu}P_{\perp}+\Delta \partial_{\mu}\log \vartheta\right).
\end{eqnarray}
At this stage it is noteworthy to mention that in deriving the above relation we have used the conservation equation (\ref{E1}) in order to obtain\footnote{In our derivation we have assumed that, $ \mathfrak{v}^{\mu}\partial_{\mu}\Delta =0 $ which corresponds to the fact that the gradient of the pressure difference is orthogonal to the anisotropy vector $ \mathfrak{v}^{\mu} $ \cite{Gahramanov:2012wz}-\cite{Florkowski:2008ag}.},
\begin{eqnarray}
\mathfrak{a}^{\mu}=\mathfrak{u}^{\nu}\partial_{\nu}\mathfrak{u}^{\mu}=\frac{1}{(\epsilon + P_{\perp})}\left( \varrho \mathfrak{E}^{\mu}-\mathfrak{P}^{\mu \nu}(\partial_{\nu}P_{\perp}-\Delta\partial_{\nu}\log \vartheta)\right). 
\end{eqnarray}
Therefore, to be precise, we are looking for an \textit{on shell} cancellation in the parity odd sector of the entropy current where the last term corresponds to the anisotropic contribution to the acceleration of the fluid particles \cite{Hoyos:2015lra}.\\ \\
\textbf{(III)}  
\begin{eqnarray}
-\frac{1}{T}\partial_{\mu}\mathfrak{u}_{\nu}\mathfrak{S}^{(\Delta , \slashed{P})\mu \nu}=(\beta_{\omega}\omega^{\mu}+\beta_{\mathfrak{B}}\mathfrak{B}^{\mu})(\mathfrak{v}^{\nu}\mathrm{Q}_{\mu \nu}-\mathfrak{v}_{\mu}\mathbb{N})
\end{eqnarray}
where, $ \mathbb{N}=\partial . \mathfrak{u}-2 \mathfrak{u}^{\mu}\partial_{\mu}\log\vartheta $.\\ \\
\textbf{(IV)}  
\begin{eqnarray}
-\mathfrak{D}^{(\Delta , \slashed{P})\mu}\left(\mathfrak{P}_{\mu \nu}\partial^{\nu}\left(\frac{\mu_{c}}{T} \right)- \frac{\mathfrak{E}_{\mu}}{T} \right)=-(\beta_{\omega}\omega^{\mu}+\beta_{\mathfrak{B}}\mathfrak{B}^{\mu})\mathfrak{v}
_{\mu}\Theta^{(P)}.
\end{eqnarray}
\textbf{(V)}  
\begin{eqnarray}
-\frac{1}{T}\mathfrak{G}^{(\Delta , \slashed{P})\nu}\mathfrak{u}^{\mu}\partial_{\mu}\mathfrak{u}_{\nu}=(\beta_{\omega}\omega^{\mu}+\beta_{\mathfrak{B}}\mathfrak{B}^{\mu})\mathfrak{v}
_{\mu}(\mathfrak{v}.\mathfrak{a}).
\end{eqnarray}
\textbf{(VI)}  
\begin{eqnarray}
\frac{1}{T}\mathfrak{u}_{\nu}\partial_{\mu}\mathfrak{Q}^{(\Delta ,\slashed{P})\mu \nu}=(\beta_{\omega}\omega^{\mu}+\beta_{\mathfrak{B}}\mathfrak{B}^{\mu})\mathfrak{v}
_{\mu}(\mathfrak{v}.\mathfrak{a}).
\end{eqnarray}

Finally, collecting all the individual pieces in (\ref{E48}) we arrive at the following set of constraint equations namely,
\begin{eqnarray}
\partial_{\mu} \mathbb{D}_{\omega}-\frac{(2 \mathbb{D}_{\omega} +\beta_{\omega})}{(\epsilon + P_{\perp})}\left(\partial_{\mu}P_{\perp}-\Delta \partial_{\mu}\log\vartheta \right)-\varsigma_{\omega}\partial_{\mu}\left(\frac{\mu_{c}}{T} \right)+\Delta \beta_{\omega}\mathfrak{v}^{\nu}(\mathrm{Q}_{\mu \nu}-g_{\mu \nu}\mathbb{Z})&=&0\nonumber\\
\frac{(2 \mathbb{D}_{\omega}+\beta_{\omega})\varrho}{(\epsilon +P_{\perp})}-2 \mathbb{D}_{\mathfrak{B}}+\frac{\varsigma_{\omega}}{T}&=&0\nonumber\\
\partial_{\mu} \mathbb{D}_{\mathfrak{B}}-\frac{(\mathbb{D}_{\mathfrak{B}} +\beta_{\mathfrak{B}})}{(\epsilon + P_{\perp})}\left(\partial_{\mu}P_{\perp}-\Delta \partial_{\mu}\log\vartheta \right)-\varsigma_{\mathfrak{B}}\partial_{\mu}\left(\frac{\mu_{c}}{T} \right)+\Delta \beta_{\mathfrak{B}}\mathfrak{v}^{\nu}(\mathrm{Q}_{\mu \nu}-g_{\mu \nu}\mathbb{Z})&=&0\nonumber\\
\frac{(\mathbb{D}_{\mathfrak{B}}+\beta_{\mathfrak{B}})\varrho}{(\epsilon +P_{\perp})}+\frac{\varsigma_{\mathfrak{B}}}{T}-\mathfrak{c}\frac{\mu_{c}}{T}&=&0
\label{E58}
\end{eqnarray}
where, the entity $ \mathbb{Z} $ could be formally expressed as,
\begin{eqnarray}
\mathbb{Z}=\mathbb{N}+\Theta^{(P)}-2 (\mathfrak{v}. \mathfrak{a}).
\end{eqnarray}

The above set of equations (\ref{E58}) precisely corresponds to the \textit{anisotropic} generalization of the earlier results \cite{Hoyos:2015lra} obtained in the context of anomalous hydrodynamics at quantum critical (Lifshitz) point. Our ultimate goal would be to solve (\ref{E58}) in order to obtain parity odd transports away from quantum critical point. In order to proceed further, we however treat the above set of equations perturbatively in the anisotropy and/ or the spin wave excitation $ (|\Delta| \ll 1 )$. We consider the following perturbative expansion of anomalous transports namely,
\begin{eqnarray}
\mathbb{Q}_{i}=\mathbb{Q}_{i}^{(0)}+\Delta \mathbb{Q}_{i}^{(1)}+\mathcal{O}(\Delta^{2})
\end{eqnarray}
where, $ \mathbb{Q}=\{\mathbb{D},\beta,\varsigma\}$ and $ i=\{\omega , \mathfrak{B}\} $. The zeroth order solutions are indeed trivial and has been found earlier in \cite{Hoyos:2015lra}. However, the first non trivial correction appears at the leading order in $ \Delta $. The corresponding set of equations at leading order turns out to be\footnote{For the sake of simplicity we do not quote the zeroth order solutions here. Interested reader may have a look at \cite{Hoyos:2015lra}. These solutions are purely fixed by the anomaly. However, we will use them explicitly in order to explore solutions at leading order in the anisotropy.},
\begin{eqnarray}
\partial_{\mu} \mathbb{D}^{(1)}_{\omega}-(2 \mathbb{D}^{(1)}_{\omega} +\beta^{(1)}_{\omega})\left( \frac{\partial_{\mu}P_{\perp}}{\epsilon + P_{\perp}}\right)  
+\frac{(2 \mathbb{D}^{(0)}_{\omega} +\beta^{(0)}_{\omega})}{(\epsilon + P_{\perp})}\partial_{\mu}\log\vartheta
-\varsigma^{(1)}_{\omega}\partial_{\mu}\left(\frac{\mu_{c}}{T} \right)+ \beta_{\omega}\mathfrak{v}^{\nu}(\mathrm{Q}_{\mu \nu}-g_{\mu \nu}\mathbb{Z})&=&0\nonumber\\
\frac{(2 \mathbb{D}^{(1)}_{\omega}+\beta^{(1)}_{\omega})\varrho}{(\epsilon +P_{\perp})}-2 \mathbb{D}^{(1)}_{\mathfrak{B}}+\frac{\varsigma^{(1)}_{\omega}}{T}&=&0\nonumber\\
\partial_{\mu} \mathbb{D}^{(1)}_{\mathfrak{B}}-( \mathbb{D}^{(1)}_{\mathfrak{B}} +\beta^{(1)}_{\mathfrak{B}})\left( \frac{\partial_{\mu}P_{\perp}}{\epsilon + P_{\perp}}\right)  
+\frac{(\mathbb{D}^{(0)}_{\mathfrak{B}} +\beta^{(0)}_{\mathfrak{B}})}{(\epsilon + P_{\perp})}\partial_{\mu}\log\vartheta
-\varsigma^{(1)}_{\mathfrak{B}}\partial_{\mu}\left(\frac{\mu_{c}}{T} \right)+ \beta_{\mathfrak{B}}\mathfrak{v}^{\nu}(\mathrm{Q}_{\mu \nu}-g_{\mu \nu}\mathbb{Z})&=&0\nonumber\\
\frac{(\mathbb{D}^{(1)}_{\mathfrak{B}}+\beta^{(1)}_{\mathfrak{B}})\varrho}{(\epsilon +P_{\perp})}+\frac{\varsigma^{(1)}_{\mathfrak{B}}}{T}&=&0.\nonumber\\
\label{E61}
\end{eqnarray}

The above set of equations (\ref{E61}) could be simplified further in order to obtain the following reduced set of equations namely,
\begin{eqnarray*}
\partial_{\mu} (\mathbb{D}^{(1)}_{\omega}-\mathbb{D}^{(1)}_{\mathfrak{B}})-\left(2\mathbb{D}^{(1)}_{\omega}
-\mathbb{D}^{(1)}_{\mathfrak{B}}+\beta_{\omega}^{(1)}-\beta_{\mathfrak{B}}^{(1)} \right) \left( \frac{\partial_{\mu}P_{\perp}}{\epsilon + P_{\perp}}\right)+\frac{\mathbb{X}^{(0)}}{(\epsilon +P_{\perp})}\partial_{\mu}\log\vartheta\nonumber\\ -(\varsigma_{\omega}^{(1)}-\varsigma_{\mathfrak{B}}^{(1)})\partial_{\mu}\left( \frac{\mu_{c}}{T}\right)=0
\end{eqnarray*}
\begin{eqnarray}
\frac{(2 \mathbb{D}^{(1)}_{\omega}+\beta^{(1)}_{\omega})\varrho}{(\epsilon +P_{\perp})}-2 \mathbb{D}^{(1)}_{\mathfrak{B}}+\frac{\varsigma^{(1)}_{\omega}}{T}&=&0\nonumber\\
\frac{(\mathbb{D}^{(1)}_{\mathfrak{B}}+\beta^{(1)}_{\mathfrak{B}})\varrho}{(\epsilon +P_{\perp})}+\frac{\varsigma^{(1)}_{\mathfrak{B}}}{T}&=&0
\label{E62}
\end{eqnarray}
where, the entity $ \mathbb{X}^{(0)} $ could be formally expressed as,
\begin{eqnarray}
\mathbb{X}^{(0)}=2\mathbb{D}^{(0)}_{\omega}
-\mathbb{D}^{(0)}_{\mathfrak{B}}+\beta_{\omega}^{(0)}-\beta_{\mathfrak{B}}^{(0)}.
\end{eqnarray}

Considering the general identity\footnote{Here we define, $ \bar{\mu}_{c}=\frac{\mu_{c}}{T} $.},
\begin{eqnarray}
\partial_{\mu}\mathbb{D}_{i}^{(1)}=\left(\frac{\partial \mathbb{D}_{i}^{(1)}}{\partial P_{\perp}} \right)_{\bar{\mu}_{c},\log \vartheta}\partial_{\mu}P_{\perp} +\left(\frac{\partial \mathbb{D}_{i}^{(1)}}{\partial \bar{\mu}_{c}} \right)_{P_{\perp} ,\log \vartheta}\partial_{\mu}\bar{\mu}_{c}+\left(\frac{\partial \mathbb{D}_{i}^{(1)}}{\partial \log \vartheta} \right)_{P_{\perp} , \bar{\mu}_{c}}\partial_{\mu}\log \vartheta
\end{eqnarray}
from (\ref{E62}), we arrive at the following set of equations,
\begin{eqnarray}
\left(\frac{\partial \mathbb{D}_{\omega}^{(1)}}{\partial P_{\perp}} \right)_{\bar{\mu}_{c},\log \vartheta}-\frac{\left(2\mathbb{D}^{(1)}_{\omega}
-\mathbb{D}^{(1)}_{\mathfrak{B}}+\beta_{\omega}^{(1)}-\beta_{\mathfrak{B}}^{(1)} \right)}{(\epsilon + P_{\perp})}&=&0\nonumber\\
\left(\frac{\partial \mathbb{D}_{\omega}^{(1)}}{\partial \bar{\mu}_{c}} \right)_{ P_{\perp},\log \vartheta}-(\varsigma_{\omega}^{(1)}-\varsigma_{\mathfrak{B}}^{(1)})&=&0
\nonumber\\
\left(\frac{\partial \mathbb{D}_{\omega}^{(1)}}{\partial \log \vartheta} \right)_{P_{\perp} , \bar{\mu}_{c}}+\frac{\mathbb{X}^{(0)}}{(\epsilon +P_{\perp})}&=&0\nonumber\\
\left(\frac{\partial \mathbb{D}_{\mathfrak{B}}^{(1)}}{\partial P_{\perp}} \right)_{\bar{\mu}_{c},\log \vartheta}+\frac{\left(2\mathbb{D}^{(1)}_{\omega}
-\mathbb{D}^{(1)}_{\mathfrak{B}}+\beta_{\omega}^{(1)}-\beta_{\mathfrak{B}}^{(1)} \right)}{(\epsilon + P_{\perp})}&=&0\nonumber\\
\left(\frac{\partial \mathbb{D}_{\mathfrak{B}}^{(1)}}{\partial \bar{\mu}_{c}} \right)_{ P_{\perp},\log \vartheta}+(\varsigma_{\omega}^{(1)}-\varsigma_{\mathfrak{B}}^{(1)})&=&0\nonumber\\
\left(\frac{\partial \mathbb{D}_{\mathfrak{B}}^{(1)}}{\partial \log \vartheta} \right)_{P_{\perp} , \bar{\mu}_{c}}-\frac{\mathbb{X}^{(0)}}{(\epsilon +P_{\perp})}&=&0\nonumber\\
\frac{(2 \mathbb{D}^{(1)}_{\omega}+\beta^{(1)}_{\omega})\varrho}{(\epsilon +P_{\perp})}-2 \mathbb{D}^{(1)}_{\mathfrak{B}}+\frac{\varsigma^{(1)}_{\omega}}{T}&=&0\nonumber\\
\frac{(\mathbb{D}^{(1)}_{\mathfrak{B}}+\beta^{(1)}_{\mathfrak{B}})\varrho}{(\epsilon +P_{\perp})}+\frac{\varsigma^{(1)}_{\mathfrak{B}}}{T}&=&0.
\label{E65}
\end{eqnarray}

Eq. (\ref{E65}) is precisely an artifact of the symmetry breaking effect which non trivially modifies the previously obtained constraint equations \cite{Hoyos:2015lra}. In order to proceed further, from (\ref{E21}) we note,
\begin{eqnarray}
\left(\frac{\partial \bar{\mu}_{c}}{\partial T} \right)_{P_{\perp} , \log \vartheta}&=&-\frac{(\epsilon +P_{\perp})}{\varrho T^{2}}\nonumber\\
\left(\frac{\partial P_{\perp}}{\partial T} \right)_{\bar{\mu}_{c} , \log \vartheta}&=&\frac{(\epsilon +P_{\perp})}{T}\nonumber\\
\left(\frac{\partial \log \vartheta}{\partial T} \right)_{\bar{\mu}_{c} , P_{\perp}}&=&-\frac{1}{\Delta}\frac{(\epsilon +P_{\perp})}{T}.
\label{E66}
\end{eqnarray}

Substituting (\ref{E66}) into (\ref{E65}) we find,
\begin{eqnarray*}
\left(\frac{\partial \mathbb{D}_{\omega}^{(1)}}{\partial \log \vartheta} \right)_{T, \bar{\mu}_{c}}+\left(\frac{\partial \mathbb{D}_{\mathfrak{B}}^{(1)}}{\partial \log \vartheta} \right)_{T, \bar{\mu}_{c}}+\mathcal{O}(\Delta^{2},T)=0
\end{eqnarray*}
\begin{eqnarray*}
\left( 1-\frac{\varrho T^{2}}{(\epsilon + P_{\perp})}\right) \left( \left(\frac{\partial \mathbb{D}_{\omega}^{(1)}}{\partial T} \right)_{\bar{\mu}_{c}, \log\vartheta}+\left(\frac{\partial \mathbb{D}_{\mathfrak{B}}^{(1)}}{\partial T} \right)_{\bar{\mu}_{c}, \log \vartheta}\right)\nonumber\\
+\left(\frac{\partial \mathbb{D}_{\omega}^{(1)}}{\partial \bar{\mu}_{c}} \right)_{T, \log\vartheta}+\left(\frac{\partial \mathbb{D}_{\mathfrak{B}}^{(1)}}{\partial \bar{\mu}_{c}} \right)_{T, \log\vartheta} =0
\end{eqnarray*}
\begin{eqnarray}
\frac{(2 \mathbb{D}^{(1)}_{\omega}+\beta^{(1)}_{\omega})\varrho}{(\epsilon +P_{\perp})}-2 \mathbb{D}^{(1)}_{\mathfrak{B}}+\frac{\varsigma^{(1)}_{\omega}}{T}&=&0\nonumber\\
\frac{(\mathbb{D}^{(1)}_{\mathfrak{B}}+\beta^{(1)}_{\mathfrak{B}})\varrho}{(\epsilon +P_{\perp})}+\frac{\varsigma^{(1)}_{\mathfrak{B}}}{T}&=&0.
\label{E67}
\end{eqnarray}

From the first two equations in (\ref{E67}), we find,
\begin{eqnarray}
\mathbb{D}^{(1)}_{\mathfrak{B}}+\mathbb{D}^{(1)}_{\omega}=\mathcal{C}
\label{E68}
\end{eqnarray}
where, $ \mathcal{C} $ is some integration constant that does not depend on any of the thermodynamic variables. Using (\ref{E68}), the last two equations of (\ref{E67}) could be further modified as,
\begin{eqnarray}
\frac{(2 \mathcal{C}+\beta^{(1)}_{\omega})\varrho}{(\epsilon +P_{\perp})}-2 \mathbb{D}^{(1)}_{\mathfrak{B}}\left(1+\frac{\varrho}{(\epsilon +P_{\perp})} \right) +\frac{\varsigma^{(1)}_{\omega}}{T}&=&0\nonumber\\
\frac{(\mathbb{D}^{(1)}_{\mathfrak{B}}+\beta^{(1)}_{\mathfrak{B}})\varrho}{(\epsilon +P_{\perp})}+\frac{\varsigma^{(1)}_{\mathfrak{B}}}{T}&=&0.
\label{E69}
\end{eqnarray}
Clearly, the above Eq.(\ref{E69}) is inadequate to solve all the five different transports in a unique fashion\footnote{Similar conclusion have been drawn earlier by other authors in \cite{Neiman:2010zi}.}. Therefore what we find from the above computation is that the entropy current cannot uniquely fix all the parity odd transports once we break the underlying symmetry of the system. However, it indeed imposes non trivial constraints among different transports as we have also noticed earlier in the parity even sector.

%%%%%%%%%%%%%%%%%%%%%%%%%%%%%%%%%%%%
\section{Summary and final remarks}
We now summarise the key findings of our analysis.
The most significant outcome of our analysis is the establishment of the hydrodynamic description away from the quantum criticality. We claim that such hydrodynamic description could be identified with the corresponding hydrodynamic description of \textit{spin waves} in the appropriate limit. The reason for this claim rests over the facts that (1) the spin wave excitation are naturally diminished as the fixed point is approached and (2) they spontaneously break the underlying rotational symmetry of the system. Both of which have been incorporated in the present analysis.  Our analysis reveals that it is indeed possible to find out new dissipative effects when we explicitly break the underlying symmetry associated with QCPs. We compute the entropy current in order to fix these arbitrary coefficients in the constitutive relation. It turns out that none of these transports could be fixed uniquely. Nevertheless, they are non trivially constrained by their respective constraint equations. We perform our analysis for the most general charged dissipative fluids where we consider both parity preserving as well as parity violating effects at leading order in the derivative expansion.  

Finally, as mentioned earlier, for realistic models of strange metal systems one should actually consider the so called \textit{non relativistic} limit of the Lifshitz hydrodynamics that has been discussed so far. This will eventually lead us towards a formulation of non relativistic QFTs with broken Galilean boost starting from a theory with broken Lorentz boost invariance. This could be done in two different ways. One could either take a $( v^{i}/c) $ limit of the theory \cite{Hoyos:2013eza} or simply by going to some light cone frame \cite{Rangamani:2008gi}-\cite{Banerjee:2014mka}. However, one should take care of the fact that in the non relativistic limit the so called scaling relations will also get modified leading to a modified Ward identity \cite{Hoyos:2013eza}. We leave all these issues for the purpose of future investigation. \\ \\
%%%%%%%%%%%%%%%%%%%%%%%%%%%%%%%%%%%%%%%%%%%%%%%%%%%%%%%%%%%%%%%%%%%%%%
{\bf {Acknowledgements :}}
 The author would like to acknowledge the financial support from UGC (Project No UGC/PHY/2014236).
 %%%%%%%%%%%%%%%%%%%%%%%%%%%%%%%%%%%%%%%%%%%%%%%%%%%%%%%%%%%%%%%%%%%%%%%%%%%%%%%%%%%

\end{document}